\documentstyle[prd,aps,preprint,psfig]{revtex}

\begin{document}
\tighten
\title{Primordial Black Hole Formation in Supergravity}
\author{M. Kawasaki}
\address{Institute for Cosmic Ray Research, University of Tokyo,
  Tanashi 188-8502, Japan}
\author{T. Yanagida} \address{Department of Physics and RESCEU,
  University of Tokyo, Tokyo 113-0033, Japan}
\date{\today}

\maketitle

\begin{abstract}
  We study a double inflation model (a preinflation $+$ a new
  inflation) in supergravity and discuss the formation of primordial black
  holes which may be identified with massive compact halo objects
  (MACHOs) observed in the halo of our galaxy.  The preinflation
  drives an inflaton for the new inflation close to the origin through
  supergravity effects and the new inflation naturally occurs.  If the
  total $e$-fold number of the new inflation is smaller than $\sim
  60$, both inflations produce cosmologically relevant density
  fluctuations.  If the coherent inflaton oscillation after the
  preinflation continues until the beginning of the new inflation,
  density fluctuations on small cosmological scales can be set
  suitably large to produce black holes MACHOs of masses $\sim 1
  M_{\odot}$ in a wide region of parameter space in the double
  inflation model.
\end{abstract}



\section{Introduction}

Massive compact halo objects (MACHOs) observed by gravitational
lensing effects~\cite{MACHO} are one of the viable candidates for dark
matter in the present universe.  The analysis of the data taken by the
MACHO collaboration~\cite{MACHO} suggests that about half of our halo
is composed of MACHOs whose masses are about $(0.5 - 0.6) M_\odot$.
It is unlikely that MACHOs are low mass stars such as red dwarfs since
the observed abundance of red dwarfs is small~\cite{Graff-Freese}.
White dwarfs may account for the MACHOs if the initial mass function
of the stellar population has a sharp peak at mass scale $\sim
2M_{\odot}$~\cite{Adams-Laughlin}.  It is a unsolved problem whether
or not such an initial mass function is consistent with observations.
Thus, there arises another possibility, that MACHOs are primordial
black holes of masses $\sim 1M_{\odot}$.  In the early universe such
black holes could be formed owing to the large density fluctuations at
cosmic temperature $T\sim 1$~GeV.

In a previous paper~\cite{KSY}, Sugiyama and the present authors
proposed\footnote{
  Different models for the primordial black hole formation have been
  studied in Ref.~\cite{Yokoyama}.}
a model for the production of black hole MACHOs using a double inflation
(a preinflation $+$ a new inflation) model in
supergravity.\footnote{
  For a review of inflation models in supergravity, e.g., see
  Ref.~\cite{Riotto}.}
This double inflation model was originally proposed to solve the
initial value problem of the new inflation model~\cite{Izawa}.  In
supergravity the reheating temperature of inflation should be low
enough to avoid overproduction of gravitinos~\cite{Ellis,Moroi}.  The
new inflation model~\cite{Albrecht} generally predicts a very low
reheating temperature and hence it is the most attractive among many
inflation models.  However, the new inflation suffers from a
fine-tuning problem about the initial condition; i.e., for successful
new inflation, the initial value of the inflaton should be very close
to the local maximum of the potential in a large region whose size is
much longer than the horizon of the universe.  It was shown that this
serious problem is solved by supergravity effects if there existed a
preinflation (e.g., hybrid inflation) with a sufficiently large Hubble
parameter before the new inflation~\cite{Izawa}.

In this double inflation model, if the total $e$-fold number of the
new inflation is smaller than $\sim 60$, density fluctuations produced
by both inflations are cosmologically relevant. In this case, the
preinflation should account for the density fluctuations on large
cosmological scales [including the Cosmic Background Explorer (COBE)
scales] while the new inflation produces density fluctuations on small
scales. Although the amplitude of the fluctuations on large scales
should be normalized to the COBE data~\cite{COBE}, fluctuations on
small scales are free from the COBE normalization and can be large
enough to produce primordial black holes which may be identified with
MACHOs.

In the previous paper~\cite{KSY}, however, it was found that the MACHO
black holes are formed in a very restricted parameter space.  This is
because large quantum fluctuations of the inflaton for the new
inflation are induced during the preinflation and they remain until
the beginning of the new inflation, which results in a too inhomogeneous
universe.  The estimation of the quantum fluctuations was based on the
assumption that the reheating takes place quickly after the
preinflation.  However, if the coherent oscillation of the inflaton
continues for a long time, the amplitude of the quantum fluctuations
of the new inflaton decreases during this oscillation phase and
becomes negligible.  Therefore, in this paper, we make a re-analysis
of the double inflation model assuming a long period of the coherent
oscillation of the (preinflation) inflaton field.  Taking a hybrid
inflation~\cite{hybrid} as an example of the preinflation, we find
that density fluctuations on small cosmological scales can have a
suitable magnitude to produce black hole MACHOs of masses $\sim 1
M_{\odot}$ in a wide region of parameter space in the double inflation
model if the coherent oscillation of an inflaton after the hybrid
inflation continues until the beginning of the new inflation.

\section{Black hole formation}

In a radiation-dominated universe, primordial black holes are formed
if the density fluctuations $\delta $ at horizon crossing satisfy a
condition $1/3 \le \delta \le 1$~\cite{Car,Green}. Masses of the black
holes $M_{BH}$ are roughly equal to the horizon mass: 
\begin{equation}
    \label{bh-mass}
    M_{BH} \simeq 4\sqrt{3}\pi\frac{M^3}{\sqrt{\rho}}
    \simeq 0.066 M_{\odot} \left(\frac{T}{{\rm GeV}}\right)^{-2},
\end{equation}
where $M (\simeq 2.4\times 10^{18}$~GeV) is the gravitational scale and
$\rho$ and $T$ are the total density and temperature of the universe,
respectively. Thus, black holes of masses $\sim 1M_{\odot}$ can be
formed at temperature $\sim 0.26$~GeV. Since we are interested in the
black holes to be identified with the MACHOs, we assume hereafter that
the temperature at the black hole formation epoch is $T_{*}\simeq
0.26$~GeV.

The horizon length at the black hole formation epoch ($T=T_*$)
corresponds to the scale $L_{*}$ in the present universe given by 
\begin{equation}
    \label{bh-scale}
    L_{*} \simeq \frac{a(T_0)}{a(T_{*})}H^{-1}(T_{*})
    \simeq 0.25~ {\rm pc},
\end{equation}
where $T_{0}$ is the temperature of the present universe.

The mass fraction $\beta_{*} (= \rho_{BH}/\rho)$ of primordial 
black holes of mass $M_{*}$ is given by~\cite{Green} 
\begin{equation}
    \label{bh-frac}
    \beta_{*}(M_{*}) = \int_{1/3}^{1} 
    \frac{d\delta}{\sqrt{2\pi}\bar{\delta}(M_*)}
    \exp\left(-\frac{\delta^2}{2\bar{\delta}^2(M_{*})}\right)
    \simeq \bar{\delta}(M_{*}) 
    \exp\left(-\frac{1}{18\bar{\delta}^2(M_{*})}\right) ,
\end{equation}
where $\bar{\delta}(M_{*})$ is the mass variance at the horizon crossing.
Assuming that only black holes of mass $M_{*}$ are formed (this
assumption is justified later), the density of the black holes
$\rho_{BH}$ is given by
\begin{equation}
    \label{bh-density}
    \frac{\rho_{BH}}{s} \simeq \frac{3}{4}\beta_{*}T_{*},
\end{equation}
where $s$ is the entropy density.  Since $\rho_{BH}/s$ is constant at 
$T<T_{*}$, we can write the density parameter $\Omega_{BH}$ of 
the black holes in the present universe as 
\begin{equation}
    \Omega_{BH}h^2 \simeq 5.6\times 10^7 \beta_{*},
\end{equation}
where we have used the present entropy density $2.9\times
10^{3}$~cm$^{-3}$ and $T_{*} \simeq 0.26$~GeV, and $h$ is the present
Hubble constant in units of 100~km/sec/Mpc.  Requiring that the black
holes ($=$MACHOs) be dark matter in the present universe, i.e.
$\Omega_{BH}h^2 \sim 0.25$, we obtain $\beta_{*} \sim 5\times 10^{-9}$
which leads to
\begin{equation}
    \bar{\delta} (M_{*}) \simeq 0.06.
\end{equation}
This mass variance suggests that the amplitude of the density
fluctuations at the mass scale $M_{*}(\simeq 1M_{\odot})$ are given by
\begin{equation}
        \frac{\delta \rho}{\rho} \simeq \frac{2}{3}\Phi  \simeq 0.01,
        \label{eq:delta-rho-BH}
\end{equation}
where $\Phi$ is the gauge-invariant fluctuations of the gravitational
potential~\cite{Linde-book}.  We will show later that such
large density fluctuations are naturally produced during the new
inflation.

\section{Double inflation model}

In this section we will consider a double inflation model in
supergravity and see how MACHO black holes are formed. We adopt the
double inflation model proposed in Refs.\cite{Izawa,KSY}.  The model
consists of two inflationary stages; the first one is called
preinflation and we adopt a hybrid inflation~\cite{hybrid} as the
preinflation. We also assume that the second inflationary stage is
realized by a new inflation model~\cite{Izawa2} and its $e$-fold number
is smaller than $\sim 60$. Thus, the density fluctuations on large
scales are produced during the preinflation and their amplitude
should be normalized by the COBE data~\cite{COBE}. On th other hand,
the new inflation produces fluctuations on small scales. Since the
amplitude of the small scale fluctuations is free from the COBE
normalization, we expect that the new inflation can produce density
fluctuations large enough to form primordial black holes.

\subsection{preinflation}

First, let us discuss a hybrid inflation model which we adopt to cause
the preinflation.  The hybrid inflation model contains two kinds of
superfields: one is $S(x,\theta)$ and the others are $\Psi(x,\theta)$
and $\bar{\Psi}(x,\theta)$. The model is also based on the U$(1)_R$
symmetry. The superpotential is given by~\cite{Copeland,hybrid}
\begin{equation}
    W(S,\Psi,\bar{\Psi}) = -\mu^{2} S + \lambda S \bar{\Psi}\Psi.
\end{equation}
The $R$-invariant K\"ahler potential is given by
\begin{equation}
    K(S,\Psi,\bar{\Psi}) = |S|^{2} + |\Psi|^{2} + |\bar{\Psi}|^{2}
    -\frac{\zeta}{4}|S|^{4} + \cdots ,
\end{equation}
where $\zeta$ is a constant of order $1$ and the ellipsis denotes
higher-order terms, which we neglect in the present analysis. We gauge
the U$(1)$ phase rotation:$\Psi \rightarrow e^{i\delta}\Psi$ and
$\bar\Psi \rightarrow e^{-i\delta}\bar\Psi$. To satisfy the $D$-term
flatness condition we take always $\Psi = \bar\Psi$ in our analysis.

Here and hereafter, we set the gravitational scale $M\simeq 2.4\times 
10^{18}$~GeV equal to unity and regard it as a plausible cutoff in 
supergravity.  As is shown in Ref.\cite{hybrid} the real part of 
$S(x)$ is identified with the inflaton field $\sigma/\sqrt{2}$.  The 
potential is minimized at $\Psi = \bar{\Psi} = 0$ for $\sigma$ larger 
than $\sigma_{c}= \sqrt{2}\mu/\sqrt{\lambda}$ and inflation occurs for 
$0< \zeta <1$ and $ \sigma_{c} \lesssim \sigma \lesssim 1$.

In a region of relatively small $\sigma$ ($\sigma_{c}\lesssim \sigma
\lesssim \lambda /\sqrt{8\pi^{2}\zeta}$) radiative corrections are
important for the inflation dynamics as shown by Dvali {\it et
  al}.\cite{Dvali}.  Including one-loop corrections, the potential for
the inflaton $\sigma$ is given by
\begin{equation}
    \label{pre-eff-pot}
    V \simeq \mu^4 \left[1 + \frac{\zeta}{2}\sigma^{2} 
    + \frac{\lambda^2}{8\pi^2}\ln\left(\frac{\sigma}{\sigma_c}\right)
    \right].
\end{equation}
The Hubble parameter $H_{\rm pre}$ and $e$-fold number $N_{\rm pre}$ 
are given by
\begin{equation}
    H_{\rm pre} \simeq \frac{\mu^{2}}{\sqrt{3}}
\end{equation}
and
\begin{equation}
    N_{\rm pre} \simeq   \left\{ \begin{array}{ll}
          \frac{1}{2\zeta} +\frac{1}{\zeta}
          \ln\frac{\sigma_{N_{\rm pre}}}{\tilde{\sigma}}
          & (\sigma_{N_{\rm pre}} >\tilde{\sigma} ),\\
          \frac{4\pi^2\sigma_{N_{\rm pre}}^2}{\lambda^{2}}
          & (\sigma_{N_{\rm pre}} < \tilde{\sigma} ),
      \end{array}\right.
    \label{Ndash}
\end{equation}
where 
\begin{equation}
    \tilde{\sigma} \simeq \frac{\lambda}{2\sqrt{2\zeta}\pi}.
\end{equation}
Here $\sigma_{N_{\rm pre}}$ is the value of the inflaton field
$\sigma$ corresponding to an $e$-fold number $N_{\rm pre}$.

If we define $N_{\rm COBE}$ as the $e$-fold number corresponding to the
COBE scale, the COBE normalization leads to a condition for the
inflaton potential:
\begin{equation}
  \label{eq:COBE-cond}
  \frac{V^{3/2}(\sigma_{\rm COBE})}{|V'(\sigma_{\rm COBE})|}
  \simeq 5.3\times 10^{-4},
\end{equation}
where $\sigma_{\rm COBE} \equiv \sigma_{N_{\rm COBE}}$.  Then, the
scale $\mu$ for the preinflation satisfies the following condition:
\begin{eqnarray}
  \label{eq:COBE-sugra}
  \frac{\mu^{2}}{\zeta\sigma_{\rm COBE}} & \simeq &
  5.3\times 10^{-4} ~~~~~~ ( \sigma_{\rm COBE} > \tilde{\sigma} ), \\[1em]
  \label{eq:COBE-one}
  \frac{8\pi^2\mu^{2}\sigma_{\rm COBE}}{\lambda^2} & \simeq &
  5.3\times 10^{-4} ~~~~~~ ( \sigma_{\rm COBE} < \tilde{\sigma}).
\end{eqnarray}
From Eqs.(\ref{Ndash}), (\ref{eq:COBE-sugra}), and (\ref{eq:COBE-one}), 
we obtain
\begin{equation}
    \label{eq:mu-N-one-loop}
    \mu \simeq 6.5\times 10^{-3} \lambda^{1/2} N_{\rm COBE}^{-1/4},
\end{equation}
for $\sigma_{\rm COBE} < \tilde{\sigma}$, and
\begin{equation}
    \label{eq:mu-N-kahler}
    \mu \simeq 6.0\times 10^{-3} \zeta^{1/4}\lambda^{1/2}
    \exp(\zeta N_{\rm COBE}/2),
\end{equation}
for $\sigma_{\rm COBE} > \tilde{\sigma}$.

\subsection{New inflation}

Now, we consider a new inflation model.  We adopt the new inflation
model proposed in Ref.~\cite{Izawa}.  The inflaton superfield $\phi(x,
\theta)$ is assumed to have an $R$ charge $2/(n+1)$ and U$(1)_{R}$ is
dynamically broken down to a discrete $Z_{2nR}$ at a scale $v$, which
generates an effective superpotential~\cite{Izawa}:
\begin{equation}
        W(\phi) = v^{2}\phi - \frac{g}{n+1}\phi^{n+1}.
        \label{sup-pot2}
\end{equation}

The $R$-invariant effective K\"ahler potential is given by 
\begin{equation}
    \label{new-kpot}
    K(\phi,\chi) = |\phi|^2 +\frac{\kappa}{4}|\phi|^4 
    + \cdots ,
\end{equation}
where $\kappa$ is a constant of order $1$.

The effective potential $V(\phi)$ for a scalar component of the
superfield $\phi(x,\theta)$ in supergravity is obtained from the above
superpotential (\ref{sup-pot2}) and the K\"ahler potential
(\ref{new-kpot}) as
\begin{equation}
        V = e^{K(\phi)}\left\{ \left(\frac{\partial^2 
        K}{\partial\phi\partial\phi^{*}}\right)^{-1}|D_{\phi}W|^{2}
        - 3 |W|^{2}\right\},
        \label{eq:new-pot}
\end{equation}
with 
\begin{equation}
        D_{\phi}W = \frac{\partial W}{\partial \phi} 
        + \frac{\partial K}{\partial \phi}W.
        \label{eq:DW}
\end{equation}
This potential yields a vacuum
\begin{equation}
    \langle \phi \rangle  \simeq  
    \left(\frac{v^2}{g}\right)^{1/n}.
\end{equation}
We have negative energy as 
\begin{equation}
        \langle V \rangle \simeq -3 e^{\langle K \rangle}
        |\langle W \rangle |^{2}
        \simeq -3 \left( \frac{n}{n+1}\right)^{2}|v|^{4}
        |\langle \phi \rangle|^{2}.
        \label{vacuum}
\end{equation}
The negative vacuum energy (\ref{vacuum}) is assumed to be canceled
out by a supersymmetry- (SUSY)-breaking effect~\cite{Izawa2} which
gives a positive contribution $\Lambda^{4}_{SUSY}$ to the vacuum
energy. Thus, we have a relation between $v$ and the gravitino mass
$m_{3/2}$:
\begin{equation}
        m_{3/2} \simeq \frac{\Lambda^{2}_{SUSY}}{\sqrt{3}}
        = \left(\frac{n}{n+1}\right) |v|^{2}
        \left|\frac{v^{2}}{g}\right|^{\frac{1}{n}}.
        \label{gravitino-mass}
\end{equation}

The inflaton $\phi$ has a mass $m_{\phi}$ in the vacuum with ( for
details, see Ref.~\cite{Izawa})
\begin{equation}
        m_{\phi} \simeq n |g|^{\frac{1}{n}}|v|^{2-\frac{2}{n}}.
        \label{inftaton-mass}
\end{equation}
The inflaton $\phi$ may decay into ordinary particles through
gravitationally suppressed interactions, which yields reheating
temperature $T_R$ given by
\begin{equation}
    \label{reheat-temp}
    T_R \simeq 0.1 m_{\phi}^{3/2} \simeq 0.1n^{\frac{3}{2}}
    |g|^{\frac{3}{2n}}|v|^{3-\frac{3}{n}}.
\end{equation}
For example, the reheating temperature $T_R$ is as low as $2
- 6 \times 10^{4}$~GeV for $v \simeq 10^{-8} - 10^{-6}$ ($m_{3/2}
\simeq 0.02~ {\rm GeV}-2$~TeV), $n=4$ and $g\simeq 1$, which is low
enough to solve the gravitino problem.\footnote{
Since the reheating temperature is low, we assume that the baryon
asymmetry is produced through the electroweak
baryogenesis~\cite{Rubakov} or the Affleck-Dine
mechanism~\cite{Affleck}.}

Let us discuss dynamics of the new inflation.  Identifying the 
inflaton field $\varphi(x)/\sqrt{2}$ with the real part of the field 
$\phi(x)$, we obtain a potential of the inflaton for $\varphi < v$
from Eq.~(\ref{eq:new-pot}):
\begin{equation}
    \label{new-eff-pot2}
    V(\varphi) \simeq v^4 - \frac{\kappa}{2}v^4\varphi^2
    -\frac{g}{2^{\frac{n}{2}-1}}v^2\varphi^n 
    + \frac{g^2}{2^n}\varphi^{2n}.
\end{equation}
It has been shown in Ref.~\cite{Izawa2} that the slow-roll condition
for the inflation is satisfied for $0< \kappa < 1$ and $\varphi
\lesssim \varphi_f$ where
\begin{equation}
    \label{new-inflaton-final}
    \varphi_f \simeq \sqrt{2}
    \left(\frac{(1-\kappa)v^2}{gn(n-1)}\right)^{\frac{1}{n-2}}.
\end{equation}
The new inflation ends when $\varphi$ becomes larger than $\varphi_f$.
The Hubble parameter of the new inflation is given by
\begin{equation}
    \label{new-hubble}
    H_{\rm new} \simeq \frac{v^2}{\sqrt{3}}.
\end{equation}
The $e$-fold number $N_{\rm new}$ is given by
\begin{equation}
    N_{\rm new} \simeq \frac{1}{\kappa} 
          \ln\left(\frac{\tilde{\varphi}}{\varphi_{N_{\rm new}}}\right)
          + \frac{1-n\kappa}{(n-2)\kappa (1-\kappa)},
    \label{eq:N-efold2}
\end{equation}
where  
\begin{equation}
    \tilde{\varphi}  =  \sqrt{2}
    \left(\frac{\kappa v^{2}}{gn}\right)^{\frac{1}{n-2}}.
    \label{eq:tilde-varphi}
\end{equation}
Here, we have assumed that $\kappa \le 1/n$. 

The amplitude of primordial density fluctuations $\delta \rho/\rho$ 
due to the new inflation is written as 
\begin{equation}
    \label{eq:new-density}
    \frac{\delta\rho}{\rho} \simeq \frac{1}{5\sqrt{3}\pi}
    \frac{V^{3/2}(\varphi_{N_{\rm new}})}{|V'(\varphi_{N_{\rm new}})|}
    = \frac{1}{5\sqrt{3}\pi} \frac{v^{2}}{\kappa\varphi_{N_{\rm new}}}.
\end{equation}
Notice here that we have larger density fluctuations for smaller
$\varphi_{N_{\rm new}}$ and hence the largest amplitude of the
fluctuations is given at the beginning of the new inflation. An
interesting point on the above density fluctuations is that it results
in a tilted spectrum with spectral index $n_s$ given
by (see Refs.~\cite{Izawa,Izawa2})
\begin{equation}
    \label{eq:new-index}
    n_s \simeq 1 - 2 \kappa.
\end{equation}
As shown later, we take $\kappa \sim 0.2$ and $n_{s}\sim 0.6$.  

\begin{figure}[htbp]
  \begin{center}
    \centerline{\psfig{figure=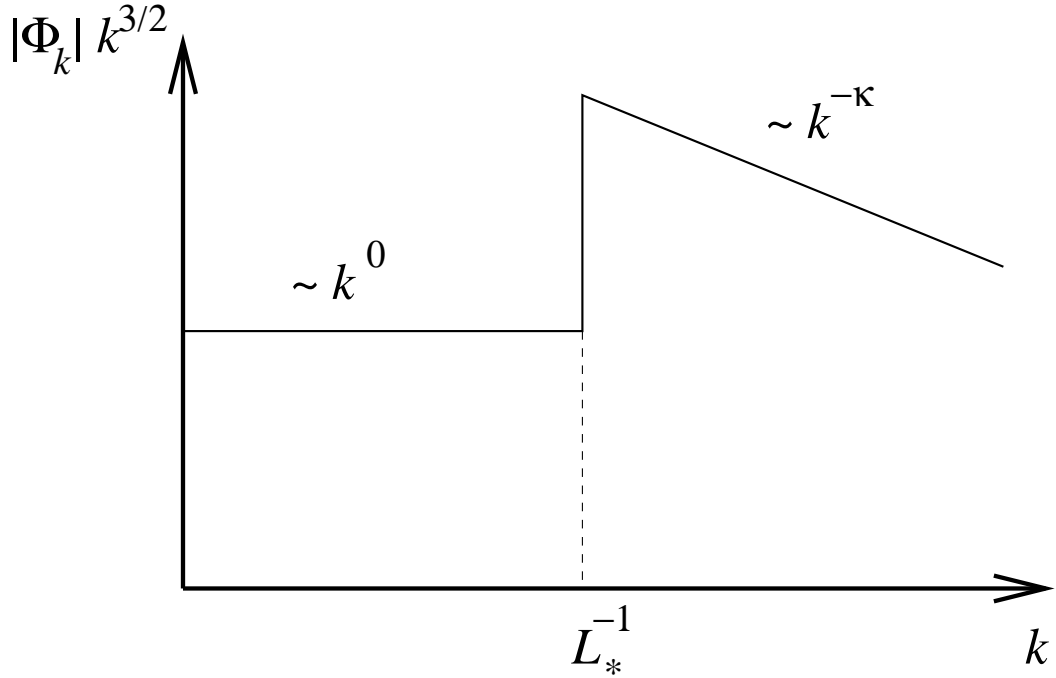,width=14cm}}
    \caption{Spectrum of the fluctuations of the gravitational
      potential $\Phi_k$ [ see Eq.~(\ref{eq:delta-rho-BH})]. $k$ is the
      comoving wave number of the fluctuations and $L_{*}$ is the
      MACHO scale. The fluctuations on large scales ($k < L_{*}^{-1}$)
      are almost scale invariant ($\sim k^{0}$) and the spectrum is
      tilted ($\sim k^{-\kappa}$) on small scales ($k > L_{*}^{-1}$).}
    \label{fig:potential}
  \end{center}
\end{figure}

Since only fluctuations produced during the new inflation have
amplitudes large enough to form the primordial black holes, the
maximum mass of the black holes is determined by fluctuations with
wavelengths equal to the horizon at the beginning of the new inflation.
We require that the maximum mass be $\sim 1M_{\odot}$ to account for
the black hole MACHOs.  On the other hand, the formation of black
holes with smaller masses is strongly suppressed since the spectrum of
the density fluctuations predicted by the new inflation is tilted [see
Eq.~(\ref{eq:new-index})]: the amplitude of the fluctuations with
smaller wavelengths is smaller (Fig.~\ref{fig:potential}).  A tiny
decrease of $\bar{\delta}(M)$ results in a large suppression of the
black hole formation rate as is seen from Eq.~(\ref{bh-frac}).
Therefore, only black holes of masses in a narrow range are formed in
the present model.

The $e$-fold number $N_{\rm new}$ is related to the present
cosmological scale $L$ by
\begin{equation}
    N_{\rm new} \simeq 60 + \ln\left(\frac{L}{3000~{\rm Mpc}}\right).
    \label{eq:N-efold}
\end{equation}
From Eq.~(\ref{eq:N-efold}), the density fluctuations corresponding to
the MACHO scale $L_{*}$ are produced when $N_{\rm new} = N_{*} \simeq
40$ during the new inflation. Since the fluctuations large enough to
produce MACHOs are induced only at the beginning of the new inflation,
$N_{*}$ is also expressed as
\begin{equation}
    N_{*} \simeq \frac{1}{\kappa} 
          \ln\left(\frac{\tilde{\varphi}}{\varphi_{b}}\right)
          + \frac{1-n\kappa}{(n-2)\kappa (1-\kappa)},
    \label{eq:Nstar-initialphi}
\end{equation}
where $\varphi_b$ is the value of $\varphi$ at the beginning
of the new inflation.

\subsection{Initial value and fluctuations of $\varphi$}

The crucial point observed in Ref.~\cite{Izawa} is that the
preinflation sets dynamically the initial condition for the new
inflation.  The inflaton field $\varphi(x)$ for the new inflation gets
an effective mass $\sim \mu^2$ from the $e^{K}[\cdots]$ term in the
potential~\cite{Copeland,Kumekawa} during the preinflation.  Thus, we
write the effective mass $m_{\rm eff}$ as
\begin{equation}
    \label{eff-mass}
    m_{\rm eff} = c\mu^2 = \sqrt{3} c H,
\end{equation}
where we introduce a free parameter $c$ since the precise value of the
effective mass depends on the details of the K\"ahler potential. For
example, if the K\"ahler potential contains $-f|\phi|^2|S|^2$, the
effective mass is equal to $\sqrt{1+f}\mu^2$.  

The evolution of the inflaton $\varphi$ for the new inflation is
described as
\begin{equation}
  \label{eq:osc-pre}
  \ddot{\varphi} + 3H \dot{\varphi} + m_{\rm eff}^2 \varphi = 0.
\end{equation}
Using $\dot{H}\simeq 0$, we get a solution to the above equation as
\begin{equation}
  \label{eq:sol-pre}
  \varphi \propto a^{-3/2 + \sqrt{9/4 - 3c^2}},
\end{equation}
where $a$ denotes the scale factor of the universe.  Thus, for $c 
\gtrsim \sqrt{3}/2$, $\varphi$ oscillates during the preinflation and 
its amplitude decreases as $a^{-3/2}$.  Thus, at the end of the 
preinflation the $\varphi$ takes a value
\begin{equation}
    \varphi \simeq \varphi_{i} 
    \exp\left(-\frac{3}{2}N_{\rm pre,tot}\right),
    \label{coherent}
\end{equation}
where $\varphi_{i}$ is the value of $\varphi$ at the beginning of the 
preinflation and $N_{\rm pre,tot}$ the total $e$-fold number of the 
preinflation.

The minimum of the potential for
$\varphi$ deviates from zero through the effect of the $|D_{S}W|^2 +
|D_{\phi}W|^2 -3|W|^2$ term  and 
this potential has a minimum as shown in Ref.~\cite{Izawa}:
\begin{equation}
    \label{deviation}
    \varphi_{\rm min} \simeq -\frac{\sqrt{2}}{c^{2}\sqrt{\lambda}}
    v\left(\frac{v}{\mu}\right).
\end{equation}
Thus, at the end of the preinflation the $\varphi$ settles down to
this $\varphi_{\rm min}$.

After the preinflation, the $\sigma$ and $\Psi (\bar\Psi)$ start to
oscillate and the universe becomes matter dominated.  $\Psi$ and
$\bar\Psi$ couple to the U$(1)$ gauge multiplets and decay immediately
to gauge fields if energetically allowed. We assume that masses for
the gauge fields are larger than those of $\Psi$ and $\bar\Psi$. We
also assume that the SUSY standard model particles do not couple to
the gauge multiplets. Thus, $S$, $\Psi$, and $\bar\Psi$ decay into
light particles only through gravitationally suppressed interactions
and the coherent oscillations of $S$, $\Psi$, and $\bar\Psi$ fields
continue until the new inflation starts.  In this period of the
coherent oscillations the average potential energy of the scalar
fields is the half of the total energy of the universe and hence the
effective mass of $\varphi$ is given by
\begin{equation}
  \label{eq:effective-mass}
  m_{\rm eff}^2  \simeq \frac{3}{2} H^2.
\end{equation}
Here and hereafter, we take $c=1$.  The evolution of $\varphi$ is
described by Eq.~(\ref{eq:osc-pre}).  Taking into account $\dot{H} =
(3/2)H^2$, one can find that the amplitude of $\varphi$ decreases as
$a^{-3/4}$.  After the preinflation ends, the superpotential for the
inflaton of the preinflation vanishes and hence the potential for
$\varphi$ has a minimum at $\varphi \simeq 0$.  Since the scale factor
increases by a factor $(\mu/v)^{4/3}$ during the matter-dominated era
between two inflations, the mean initial value $\varphi_b$ of
$\varphi$ at the
beginning of the new inflation is written as\footnote{
  Since Eq.(\ref{eq:init-new-inflaton}) represents the amplitude of
  the oscillating $\varphi$, the actual value of $\varphi_b$ should be
  multiplied by a factor $0 \le \xi \le 1$. Here, we take $\xi =1$ for
  simplicity.}
\begin{equation}
    \label{eq:init-new-inflaton}
    \varphi_b \simeq 
    \frac{\sqrt{2}}{\sqrt{\lambda}}
    v\left(\frac{v}{\mu}\right)^2.
\end{equation}

We now discuss quantum effects during the preinflation.  It is known
that in  a de Sitter universe massless fields have quantum
fluctuations whose amplitudes are given by $H/(2\pi)$.  However, the
quantum fluctuations for $\varphi$ are strongly
suppressed~\cite{Enquvist} in the present model since the mass of
$\varphi$ is larger than the Hubble parameter until the start of the new
inflation.

Let us consider the amplitude of fluctuations with comoving wavelength
$\ell_{b}$ corresponding to the horizon scale at the beginning of the
new inflation.  These fluctuations are induced during the preinflation
and its amplitude at horizon crossing [$\ell_{b} a(t_{h} = H_{\rm
pre}$ is given by $H_{\rm pre}/(2\pi)(H_{\rm pre}/m_{\rm
eff})^{1/2}$].  Since those fluctuations reenter the horizon at the
beginning of the new inflation ($t=t_b$), the scale factor of the
universe increases from $t_{h}$ to $t_b$ by a factor of $(H_{\rm
pre}/H_{\rm new})= (\mu/v)^2$.  The amplitude of fluctuations
decreases as $a^{-3/2}$ during the preinflation and $a^{-3/4}$ during
the matter-dominated era between two inflations, and the amplitude of
fluctuations with comoving wavelength $\ell_{b}$ corresponding to the
horizon scale at the beginning of the new inflation is now given by
\begin{equation}
     \delta \varphi \simeq
     \frac{H}{2\pi}
     \left(\frac{H}{m_{\rm eff}}\right)^{\frac{1}{2}}
     (\mu/v)^{(3/2)(2-4/3)} (\mu/v)^{(3/4)(4/3)}
     \simeq \frac{H}{3^{1/4}2\pi}
     \left(\frac{v}{\mu}\right)^{2}.
     \label{eq:q-fluctuation}
\end{equation}
Here, we have used the fact that the scale factor increases by
$(\mu/v)^{4/3}$ during the matter-dominated era.  The fluctuations
given by Eq.~(\ref{eq:q-fluctuation}) are a little less than newly
induced fluctuations at the beginning of the new inflation [$\simeq
v^2/(2\pi\sqrt{3}$)]. Thus, we assume that the fluctuations of
$\varphi$ induced in the preinflation can be neglected when we
estimate the fluctuations during the new inflation.

Finally, we make a comment on the domain wall problem in the double
inflation model. Since the potential of the inflaton $\phi$ has a
discrete symmetry [see Eqs.~(\ref{sup-pot2}) and (\ref{new-kpot})],
domain walls are produced if the phases of $\phi$ are spatially
random. However, the preinflation make the phase of $\phi$
homogeneous with the help of the interactions between two inflaton fields
$S$ and $\phi$ [see Eq.~(\ref{deviation})]. Therefore, the domain wall
problem does not exist in the present model.

\section{MACHO formation}

Since the density fluctuations corresponding to $L_{*}$ are produced
at the beginning of the new inflation ($N_{\rm new} = N_{*}$), from
Eqs.~(\ref{eq:delta-rho-BH}), (\ref{eq:new-density}) and
(\ref{eq:Nstar-initialphi}) we obtain
\begin{equation}
    \label{eq:new-density2}
    \frac{V^{3/2}}{V} \simeq \frac{v^2}{\kappa \varphi_b} \simeq 
    0.3,
\end{equation}
where $\varphi_b$ is given by Eq.(\ref{eq:init-new-inflaton}) and we 
have used Eq.(\ref{eq:delta-rho-BH}).  Eqs.(\ref{eq:tilde-varphi}) and 
(\ref{eq:Nstar-initialphi}) lead to 
\begin{equation}
    \label{eq:v-k}
    v \simeq 0.3 \kappa \exp(-\kappa N_{*}),
\end{equation}
where we have taken $n=4$ and $\sqrt{\kappa/(2g)}\simeq 1$, and
neglected the second term on the right-hand side of
Eq.~(\ref{eq:Nstar-initialphi}).

On the other hand, the density fluctuations produced in the
preinflation should be normalized by the COBE data, which determine
the scale of the preinflation as a function of $N_{\rm COBE}$ as
Eqs.(\ref{eq:COBE-cond})--(\ref{eq:mu-N-kahler}).  In estimating
$N_{\rm COBE}$ we must take into account the fact that fluctuations
induced at $e$-fold numbers less than $(2/3)\ln(\mu/v)$  reenter the
horizon before the new inflation starts.  Such fluctuations
are cosmologically irrelevant since the new inflation produces much
larger fluctuations.  Thus, $N_{\rm COBE}$ is given by
\begin{equation}
    \label{eq:N-COBE}
    N_{\rm COBE} = 60-N_{*} + \frac{2}{3}\ln\left(\frac{\mu}{v}\right).
\end{equation}

From Eqs.~(\ref{eq:mu-N-one-loop}), (\ref{eq:mu-N-kahler}), and
(\ref{eq:new-density2}), the scale of the new inflation is written as
\begin{equation}
  \label{eq:new-scale-one}
  v \simeq 9.95\times 10^{-5}\kappa^{-1} 
  N_{\rm COBE}^{-1/2}\lambda^{3/2},
\end{equation}
for small $\zeta$ less than $0.021$, which comes from the condition
that the one-loop corrections govern the preinflation dynamics, i.e.,
$\sigma_{N_{\rm COBE}}\equiv \sigma_{\rm COBE} < \tilde{\sigma}$.  On
the other hand, for $\zeta > 0.021$, we obtain
\begin{equation}
  \label{eq:new-scale-kahler}
  v \simeq 8.53 \times 10^{-5}\kappa^{-1}\lambda^{3/2}
  \zeta^{1/2}\exp(\zeta N_{\rm COBE}).
\end{equation}
From Eqs.(\ref{eq:N-COBE}), (\ref{eq:mu-N-one-loop}),
(\ref{eq:mu-N-kahler}), (\ref{eq:new-scale-one}), and
(\ref{eq:new-scale-kahler}) $N_{\rm COBE}$ is approximately given by
\begin{equation}
  \label{eq:N-COBE-final}
  N_{\rm COBE} \simeq 23,
\end{equation}
for both small and large $\zeta$.  Here, we have neglected ${\cal 
O}(\ln\lambda), {\cal O}(\ln\kappa)$, and ${\cal O}(\ln\zeta)$ 
corrections.  Then, we obtain $\mu$, $v$, and $\sigma_{\rm COBE}$ for 
$\zeta < 0.021$ as
\begin{eqnarray}
  \label{eq:mu-final}
  \mu & \simeq & 3.0\times 10^{-3}\lambda^{1/2},\\
  v & \simeq & 2.1 \times 10^{-5}\kappa^{-1} \lambda^{3/2},\\
  \label{eq:sigma-COBE-final}
  \sigma_{\rm COBE} & \simeq & 0.77\lambda.
\end{eqnarray}
From Eq.~(\ref{eq:v-k}), $\kappa$ is given by 
\begin{equation}
  \label{eq:kappa-final}
  \kappa \simeq 0.24,
\end{equation}
which results in
\begin{equation}
  \label{eq:v-final}
  v \simeq 8.6\times 10^{-5}\lambda^{3/2}.
\end{equation}

The coupling $\lambda$ cannot be taken too small because $\sigma_{\rm
  COBE}$ becomes less than $\sigma_c$, which leads to the constraint
\begin{equation}
  \label{eq:lambda-min}
  \lambda \gtrsim \lambda_{\rm min} = 5.5\times 10^{-3}.
\end{equation}
In the present model, the scale $v$ of the new inflation is related to
the gravitino mass as Eq.~(\ref{gravitino-mass}). Thus, the lower limit
to $\lambda$ predicts a lower bound of the gravitino mass:
\begin{equation}
  \label{eq:lower-gravitiono}
  m_{3/2} \gtrsim m_{3/2,{\rm min}} = 0.42~{\rm GeV},
\end{equation}
which is consistent with both gauge-mediated~\cite{Giudice} and
gravity mediated~\cite{Nilles} SUSY breaking models. If we require
that the gravitino mass should be less than about 1~TeV, we obtain the
upper limit to $\lambda$ given by\footnote{
  If we disconnect the scale $v$ from the gravitino mass $m_{3/2}$
  assuming some other contribution to cancel the energy in $W$ in
  Eq.~(\ref{vacuum}), this constraint may be relaxed.}
\begin{equation}
  \label{eq:lambda-max}
  \lambda \lesssim \lambda_{\rm max} = 4.3\times 10^{-2}.
\end{equation}

For the case of large $\zeta$, if we fix $\zeta$, we can obtain $\mu$,
$v$, $\kappa$, and $\sigma_{\rm COBE}$ in the same way as for the case of
small $\zeta$. The result is shown in Table~\ref{table:kahler} for
$\zeta = 0.1, 0.05$, and $0.025$. In this case, lower limit
$\lambda_{\rm min}$ is obtained by requiring $\tilde{\sigma} \gtrsim
\sigma_c$, and $m_{3/2,{\rm min}}$ and $\lambda_{\rm max}$ are
determined in the same way as in the previous case. These limits are also
shown in Table~\ref{table:kahler}. We require that $\lambda_{\rm min}
\le \lambda_{\rm max}$ for consistency.  From Table~\ref{table:kahler}
one can see that the present model works for $\zeta \lesssim 0.05$.

\section{Conclusion}

In this paper we have studied the formation of primordial black holes
of masses about $1M_{\odot}$ by taking a double inflation model in
supergravity.  We have shown that in a wide range of parameter space
primordial black holes are produced of masses $\sim 1 M_{\odot}$
which may be identified with MACHOs in the halo of our galaxy.  It
should be noticed that the present model allows natural values for the
coefficient $c$ of the effective mass of $\varphi$ during the
preinflation [i.e., $c={\cal O}(1)$].  In contrast, we must assume
unnaturally large $c$~\cite{KSY} if the inflaton decays quickly after
the preinflation.

Our double inflation model consists of a preinflation (e.g., hybrid
inflation ) and a new inflation.  The preinflation provides the
density fluctuations observed by COBE and it also dynamically sets the
initial condition of the new inflation through supergravity effects.
The predicted power spectrum has almost a scale invariant form ($n_{s}
\simeq 1$) on large cosmological scales which is favored for the
structure formation of the universe~\cite{White}.  On the other hand,
the new inflation gives the power spectrum which has large amplitude
and shallow slope ($n_{s} < 1$) on small scales.  Thus, this power
spectrum has a large and sharp peak on the scale corresponding to the
turning epoch from preinflation to new inflation.  With taking the
MACHO scale ($\sim 0.25$~pc) as the turning scale, our model leads to
the formation of the black hole MACHOs in a narrow mass range, which
is quite consistent with the observation by the MACHO
collaboration~\cite{MACHO}. We should stress that although we have
adopted a specific model for the preinflation and the new inflation,
our result is quite general in double inflation models which consist
of a generic preinflation and a new inflation. We may construct a
larger class of double inflation models which explain naturally
primordial black hole MACHOs.

In the present paper we set the turning epoch from one inflation to
another so that the masses of black holes are $\sim 1M_{\odot}$ to
account for the MACHOs. However, the double inflation model can
produce black holes of different masses by changing the $e$-fold
number of the new inflation. One of interesting cases is that produced
black holes have masses of $\sim 10^{-19}M_{\odot}$ and are just
evaporating now. Such black holes are one of the interesting
candidates for the sources of antiproton fluxes recently observed in the
BESS detector~\cite{BESS}.

The primordial black holes play the role of the usual cold dark matter on
the large scale structure formation.  The scales of the fluctuations
for primordial black hole formation themselves are much smaller
than the galactic scale and thus we cannot see any signals for such
fluctuations in $\delta T/T$ measurements.  The primordial black holes
are also attractive as a source of gravitational waves.  If the
primordial black holes dominate dark matter of the present universe,
some of them likely form binaries.  Such binary black holes coalesce
and produce significant gravitational waves~\cite{Nakamura} which may
be detectable in future detectors.

\section*{Acknowledgment}
M.K. thanks S. Kasuya and H. Suzuki for useful discussions. We also
thank S. Olito for information on the BESS results.

\begin{table}[h]
  \begin{center}
     \caption{$\mu$, $v$,  $\kappa$, $\sigma_{\rm COBE}$, $m_{3/2,{\rm
           min}}$ $\lambda_{\rm max}$ and $\lambda_{\rm min}$ for
         $\zeta=0.1, 0.05, 0.025$}
     \label{table:kahler}
     \begin{tabular}{|c|ccc|}
        $\zeta$ & $0.1$  & $0.05$ &  $0.025$  \\
        \hline
        $\mu$   
                & $1.1\times 10^{-2}\lambda^{1/2}$ 
                & $5.1\times 10^{-3}\lambda^{1/2}$ 
                & $3.2\times 10^{-3}\lambda^{1/2}$ \\
        $v$
                & $1.5\times 10^{-3}\lambda^{3/2}$
                & $2.8\times 10^{-4}\lambda^{3/2}$ 
                & $1.0\times 10^{-4}\lambda^{3/2}$ \\
        $\kappa$     
                & $0.18$ 
                & $0.21$ 
                & $0.24$ \\
        $\sigma_{\rm COBE}$
                & $2.2 \lambda$ 
                & $0.96 \lambda$ 
                & $0.77 \lambda$ \\
        $m_{3/2,{\rm min}}$     
                & $1.3\times 10^{6}~ {\rm GeV}$ 
                & $3.0\times 10^{2}~ {\rm GeV}$ 
                & $1.1~  {\rm GeV}$ \\
        $\lambda_{\rm max}$     
                & $6.3\times 10^{-3}$ 
                & $2.0\times 10^{-2}$ 
                & $3.9\times 10^{-2}$ \\
        $\lambda_{\rm min}$     
                & $4.2\times 10^{-2}$ 
                & $1.4\times 10^{-2}$ 
                & $6.3\times 10^{-3}$ \\
    \end{tabular}
  \end{center}
\end{table}

\end{document}